\documentclass{appolb}
\usepackage{graphicx}
\usepackage{xcolor}
\usepackage{cite}
\usepackage{amsmath,amssymb}
\usepackage{mathrsfs,wasysym}
\usepackage[T1]{fontenc} 
\usepackage{lmodern} 
\usepackage{hyperref}  

\newcommand{\as}{\alpha_s}
\newcommand{\Lum}{\mathscr{L}}

\newcommand{\calC}{\mathcal{C}}
\newcommand{\calL}{\mathcal{L}}
\newcommand{\calF}{\mathcal{F}}
\newcommand{\bbar}{\bar{b}}
\newcommand{\hell}{\href{https://www.roma1.infn.it/~bonvini/hell/}{\texttt{HELL}}}
\begin{document}
\title{Small-$x$ resummation in coefficient function for differential heavy-quarks production%
\thanks{Presented at ``Diffraction and Low-$x$ 2022'', Corigliano Calabro (Italy), September 24-30, 2022.}%
\author{Federico Silvetti}
\address{Dipartimento di Fisica, Sapienza Universit\`a di Roma,\\ Piazzale Aldo Moro~5, 00185 Roma, Italy}
\address{INFN, Sezione Roma 1,\\ Piazzale Aldo Moro~5, 00185 Roma, Italy}
}
\maketitle
\begin{abstract}
  \noindent  High-energy logarithmic corrections are enhanced when the ratio $x=\frac{Q^2}{s}$ between the typical energy scale of a scattering process $Q$ and the total center-of-mass energy available $s$ is small.  We discuss recent developments on their resummation in differential cross sections in rapidity, transverse momentum, and invariant mass and their application to heavy-flavor production at the LHC.
\end{abstract}
  
\section{Introduction}
\noindent Within the scope of precision physics at the LHC, many different theory computations are required to achieve an accuracy of 1\% and below (e.g. high-precision fixed-order computations, accurate PDFs and all-order resummation of large logarithmic contributions).
In this proceeding, we take into account the so-called high-energy logarithms of the form
$\as^n\frac1x\log^k\frac1x$, $k<n$, where $x = \frac{Q^2}{s}$ is a dimensionless scaling variable that becomes small when the collider energy $s$ is larger than the characteristic energy scale of a scattering process $Q$.
Then, when $x$ is small enough, the presence of such terms invalidates the perturbativity of fixed-order predictions in both the DGLAP splitting functions and the partonic cross sections.
In the former case, the theory to control high-energy logarithms has been developed over the last thirty years~\cite{Lipatov:1976zz,Fadin:1998py,Salam:1998tj,Ciafaloni:1999yw,Ciafaloni:2007gf,Ball:1995vc,Altarelli:2008aj,Thorne:1999sg,Ball:2017otu,Abdolmaleki:2018jln,Bonvini:2019wxf}.
In the latter one, the resummation of these logarithms was developed to leading logarithmic (LL) order using the $k_t$-factorization theorem~\cite{Catani:1990xk,Catani:1990eg,Catani:1994sq}. More recently, this technique received a stable numerical implementation in the High-Energy Large Logarithms public code (\hell)~\cite{Bonvini:2016wki,Bonvini:2017ogt,Bonvini:2018iwt}.

In this work, we consider the resummation of small-$x$ logarithms in cross sections differential in invariant mass, rapidity and transverse momentum. This was first developed in Refs.~\cite{Caola:2010kv,Forte:2015gve,Muselli:2017fuf}.
Here, instead, we apply the modern resummation formalism of Refs.~\cite{Bonvini:2016wki,Bonvini:2017ogt,Bonvini:2018iwt}.
Then, leveraging this numerical implementation, we consider heavy-flavor pair production as a case study and construct resummed predictions for distribution in invariant mass, transverse momentum and rapidity of the heavy-quark pair or one of the heavy quarks.

\section{Small-$x$ resummation for differential coefficient functions}
In this section, we will briefly review the resummation formalism implemented in \hell for differential observables, a more detailed discussion can be found in ~\cite{Silvetti:2022hyc}.
First, we may consider the expression for a triple-differential cross-section in the usual collinear factorization
\begin{align}
 &\frac{d \sigma}{d Q^2 d Y dq_{t}^{2}}(x,Y,Q^2,q_t^2)
  \nonumber \\
  & \qquad = \int_x^1\frac{dz}{z} \int d\hat{y} \, \Lum_{ij} \left( \frac{x}{ z},\hat y,Q^2\right) \frac{dC_{ij}}{dQ^2dydq_t^2}(z,Y-\hat y,Q^2,q_t^2)\,, \label{eq:collfact}\\
  &\Lum_{ij}\left(x,\hat y,Q^2\right) =  f_i(\sqrt{x}e^{\hat y},Q^2)\, f_j(\sqrt{x}e^{-\hat y},Q^2)\, \theta(e^{-2|\hat y|}-x)\,, \label{eq:lumicol}
\end{align}
where the sum $i,j$ runs over all partons, $Q^2,\,Y$ and $q_t^2$ are respectively the invariant mass, rapidity and transverse momentum of the final state. Inside the convolution, the parton level coefficient function $C$ is computed in the center-of-mass frame of the initial state partons and then weighted against the parton luminosity $\Lum$ of eq. \eqref{eq:lumicol}. This second object contains the usual collinear PDFs and enforces the suitable boundaries on the rapidity integral.
The same process can be written in  $k_t$-factorization ~\cite{Caola:2010kv,Forte:2015gve,Muselli:2017fuf}
 \begin{align}
   &\frac{d \sigma}{d Q^2 d Y dq_{t}^{2}}(x,Y,Q^2,q_t^2)  = \int_x^1\frac{dz}{z} \int d\hat y \int_0^\infty dk_1^2 \int_0^\infty dk_2^2\, \calL \left( \frac x z,\hat y,k_1^2,k_2^2\right) \nonumber \\
   &\qquad \times  \frac{d{\calC}}{dQ^2dydq_t^2}(z,Y-\hat y,Q^2,q_t^2, k_1^2, k_2^2)\,, \label{eq:ktfact}\\
   &{\calL}\left(x,\hat y,k_1^2,k_2^2\right) =  {\calF} (\sqrt{x}e^{\hat y},k_1^2)\, {\calF} (\sqrt{x}e^{-\hat y},k_2^2)\,, \theta(e^{-2|\hat y|}-x)\,,
 \end{align}
 where this time the coefficient function is computed at Born-level with off-shell gluons in the initial state, according to the prescription of ~\cite{Catani:1990xk,Catani:1990eg,Catani:1994sq}. Since high-energy logarithms emerge from the integration over gluon exchanges in the $t$ channel, this ensures that the LL contributions are hidden away in the off-shell luminosity $\calL$ \cite{Catani:1994sq,Caola:2010kv}. In turn, the off-shell gluon distributions $\calF$ within $\calL$ are related to their on-shell counterparts by an evolution operator $U$, which encodes the resummed logarithms and is available from Refs. ~\cite{Bonvini:2016wki,Bonvini:2017ogt,Bonvini:2018iwt,Silvetti:2022hyc}. 
This relation can be made explicit after taking a Mellin-Fourier transform over the variables $x$ and $Y$\footnote{For briefness, we write explicitly only the gluon part of the evolution relation, leaving the discussion of the quark side to a more complete review of the topic~\cite{Silvetti:2022hyc}.},
 \begin{equation}
  \calF \left( N \pm  \frac{ib}{2}, k_t^2\right) = \frac{dU}{dk_t^2}\left(N\pm \frac{ib}{2}, k_t^2, Q^2\right) f_{g}\left(N\pm \frac{ib}{2}, Q^2\right)\, . \label{eq:Fdefinition}
 \end{equation}
 Then, in this adjoint space, we can compare directly eqs. \eqref{eq:collfact} and \eqref{eq:ktfact}. Plugging in definition \eqref{eq:Fdefinition}, we can then see that
  \begin{align}
    &\frac{d C}{dQ^2dydq_t^2}(N,b,Q^2,q_t^2)  \nonumber \\
    & \qquad=\int_0^\infty dk_1^2 \int_0^\infty dk_2^2 \,  \frac{d}{dk_1^2}U\left(N+\frac{ib}2,k_1^2,Q^2\right) \nonumber \\
  &\qquad \times \frac{d}{dk_1^2}U\left(N-\frac{ib}2,k_2^2,Q^2\right) \frac{d\calC}{dQ^2dydq_t^2}(N,b,Q^2,q_t^2, k_1^2, k_2^2)\; ,
  \end{align}
  this last expression still has the structure of a Mellin-Fourier convolution like eq. \eqref{eq:collfact}. Then, after undoing the transformation, we finally retrieve
 \begin{align}
 &\frac{dC}{dQ^2dydq_t^2}(x,y,Q^2,q_t^2) \nonumber \\
 &\qquad =\int_0^\infty dk_1^2 \int_0^\infty dk_2^2 \int_x^1\frac{dz}{z} \int d\hat y\,
     \frac{d{\calC}}{dQ^2 dy dq_t^2}(z,y-\hat y,Q^2, k_1^2, k_2^2)\\
   &\qquad \times  \frac{d}{dk_1^2}U\left(\sqrt{\frac xz}e^{\hat y},k_1^2,Q^2\right) \frac{d}{dk_2^2}U\left(\sqrt{\frac xz}e^{-\hat y},k_2^2,Q^2\right)
     \, \theta\left(e^{-2|\hat y|}-\frac xz\right)\, . \label{eq:resummed}
 \end{align}
  This direct-space expression is implemented for numerical integration in the \hell ~ public code and provides a handle on the resummation for the triple-differential distribution.

   \subsection{Heavy-quark pair production}
   
\begin{figure}[t]
  \centering
  \includegraphics[width=0.49\textwidth,page=3]{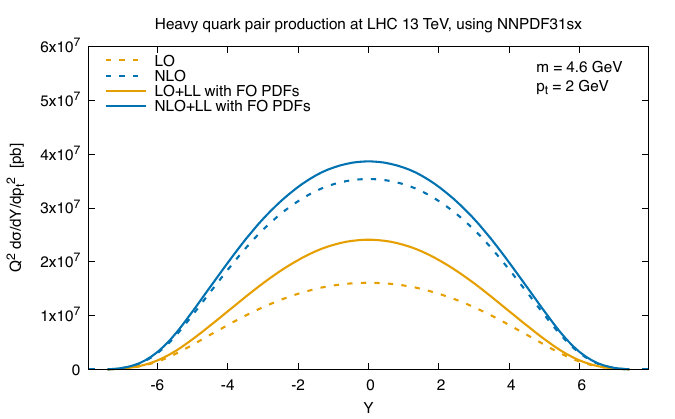}
  \includegraphics[width=0.49\textwidth,page=5]{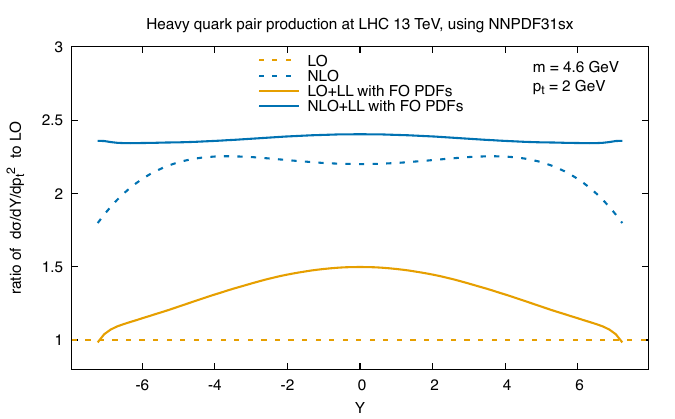}
  \caption{{\small Distribution in rapidity and transverse momentum for the bottom quark kinematics. Plotted for $q_t=2$~GeV and at LHC $13$~TeV and the NNPDF31sx PDF set.}}
  \label{fig:SQDoubleDiff}
\end{figure}
\noindent   After elucidating the structure of the new resummation formalism, we showcase some examples. We consider heavy-quark pair production (bottom flavor specifically) as an explicit case study:
\begin{equation}
  p \bar p \rightarrow b(p_b) \bbar (p_{\bbar} ) + X\; .
\end{equation}
We plot $\frac{d\sigma}{dQ^2dYdq_t^2}$ for two different kinematical settings: considering the kinematics of a single open quark or those of the entire quark-antiquark pair. In both cases, we combine the resummed result from eq. \eqref{eq:resummed} with PDF from NNPDF31sx and match to the LO and NLO fixed orders obtained from POWHEG-box~\cite{Nason:2004rx,Frixione:2007vw,Alioli:2010xd,Nason:1987xz,Frixione:2007nw}.
In the first case, we identify $Q^2$ with the squared mass of the bottom quark and the triple distribution reduces to a double one in $ q_t^2 = \left(p_{b,1} \right)^2 + \left(p_{b,2} \right)^2,\; Y = \frac{1}{2}\log\left(\frac{\left(p_{b,0} +p_{b,3} \right)}{\left(p_{b,0} - p_{\bbar,3} \right)}\right),$ reported in Fig. \ref{fig:SQDoubleDiff}.
We observe that resummation is a positive correction at LO, of about 50\% at central rapidity
and decreasing towards the rapidity endpoints.
At NLO, the correction is smaller, hinting at an improved convergence of perturbation theory from the inclusion of resummation.
Overall, the matched NLO+LL curve is about 1.4 times the LO one.  
The NLO correction alone  induces most of this effect but the contribution from the resummation grows up to 50\% at large rapidity.
\begin{figure}[t]
  \centering
  \includegraphics[width=0.49\textwidth,page=3]{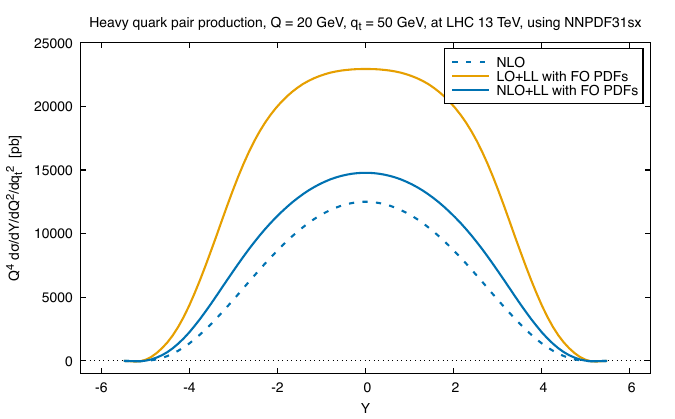}
  \caption{{\small The triple-differential distribution for the bottom-antibottom pair,
      plotted at $Q=20$~GeV and $q_t=50$~GeV, with LHC energy of $13$~TeV and the NNPDF31sx PDF set.}}
  \label{fig:TripleDiff}
\end{figure}

When considering pair kinematics instead, the two heavy bottom quarks are treated as a single object and we can identify
\begin{align*}
  & Q^2 = \left(p_b + p_{\bbar} \right)^2 ,\qquad q_t^2 = \left(p_{b,1} + p_{\bbar ,1} \right)^2 + \left(p_{b,2} + p_{\bbar, 2} \right)^2 \, ,\\
  & Y = \frac{1}{2}\log\left(\frac{\left(p_{b,0} + p_{\bbar,0} \right)+\left(p_{b,3} + p_{\bbar,3} \right)}{\left(p_{b,0} + p_{\bbar,0} \right)-\left(p_{b,3} + p_{\bbar,3} \right)}\right) \; .
\end{align*}
The corresponding plot is shown in Fig. \ref{fig:TripleDiff}.
We see that NLO (blue dashed curve) is smaller than the LL\footnote{effectively LO+LL, since the LO contribution is proportional to $\delta(q_t^2)$ and so vanishes for any non-zero transverse momentum)} curve (solid orange) result.
This time the matched NLO+LL curve (solid blue) is a small positive correction to the NLO result, pointing toward the still larger LL prediction, possibly indicating that the addition of resummation effects can bridge the gap between fixed orders and lead to a more stable perturbative expansion. It must be pointed out, however, that this apparent stabilization may be sensitive to further input from fixed order ~\cite{Ablinger:2022wbb}. 

\bibliographystyle{jhep}
\bibliography{references}

\end{document}